\documentclass{aastex}
\usepackage{amssymb,amsmath}
\usepackage{lineno}
\linenumbers
\def\app#1#2{%
	\mathrel{%
		\setbox0=\hbox{$#1\approx$}%
		\setbox2=\hbox{%
			\rlap{\hbox{$#1\propto$}}%
			\lower1.1\ht0\box0%
			}%
			\raise0.25\ht2\box2%
			}%
			}

\usepackage[hyphens]{url}
\usepackage{revsymb}
\nolinenumbers
\begin{document}
\title{Collimation of Fast Radio Burster 20201124A; Repeaters {\it vs.\/}
Apparent Non-Repeaters}
\shorttitle{Collimation of FRB 20201124A}
\shortauthors{Katz}
\author{J. I. Katz\altaffilmark{}}
\affil{Department of Physics and McDonnell Center for the Space Sciences,
Washington University, St. Louis, Mo. 63130}
\email{katz@wuphys.wustl.edu}
\begin{abstract}
	The recent report of a period in the active repeating Fast Radio
	Burster 20201124A and of its spindown rate place bounds on the solid
	angle of its emission on the basis of energetics.  The bound depends
	on the (unknown) efficiency of conversion of rotational energy to
	coherent radio emission and implies a lower bound on the Lorentz
	factor of the radiating charges.  Bursts may be emitted along the
	magnetic dipole axis, in repeaters aligned with the rotational axis
	and the line of sight but misaligned in apparent non-repeaters.
	This may explain the difficulty of finding periodicity in repeaters
	and the low duty cycle of apparent non-repeaters.
\end{abstract}
\keywords{Magnetars}
\newpage
\section{Introduction}
Fast Radio Burst (FRB) energetics remain enigmatic.  The report \citep{D25}
of a 1.7 s period during two observations, 37 days apart, of the very active
repeating FRB 20201124A implied that at least this FRB is produced by a
rotating magnetic neutron star, and the similarity of the two reported
periods supported their reality.  The statistical significance of these
detections was recently challenged \citep{GZ25}.  \citet{Z25} reported a
correlation between the the times of the claimed periodicities and changes
in Rotation Measure, which may support the reality of the periodicities.
The present paper, a preliminary version of which was published \citep{K25}
before that challenge, discusses inferences that may be drawn if periodicity
and spindown are confirmed, either for FRB 20201124A or for some other
repeating FRB.

A third epoch period determination is
required to exclude (or demonstrate) that the change in period is the result
of orbital motion.  If the small difference (about one part in 1000) of the
measured periods is attributed to spindown, the spindown age is 44 years,
implying a very young object.  The inferred surface magnetic field is $\sim
10^{15}$ Gauss; such objects are called magnetars.  Ockham's razor then
suggests that all repeating FRB are produced by magnetars.

This Research Note assumes that the bursts of FRB 20201124A are powered by
spindown rather than magnetostatic energy, that FRB have the same mechanism
as radio pulsars (PSR) rather than soft gamma repeaters (SGR): FRB resemble
PSR in emitting brief bursts of coherent radio radiation, sometimes with
bandwidth-duration product approaching the uncertainty principle limit
\citep{K24a}, rather than the smoothly varying thermal gamma-rays of SGR.
\section{Energetics}
If there is no energy reservoir intermediate between rotational energy and
radiated energy, an assumption justified by the $\sim 30\,\mu$s relaxation
time of a neutron star magnetosphere, then comparing the spindown power
inferred from the period derivative to the observed flux during bursts sets
an upper limit to the solid angle into which the bursts are emitted.  This
implies a lower bound on the Lorentz factors of the radiating charge
bunches.  It also implies an upper bound on the angular divergence of the
field lines on which the radiating charges move, and hence on the size of
the radiating region.  This last bound is unrelated to the angular width of
radiation of individual charge bunches.

The spindown power of FRB 20201124A
\begin{equation}
	\label{spindown}
	P_{spindown} = I \omega {\dot\omega} \approx 2 \times 10^{36} I_{45}
	\ \text{erg/s},
\end{equation}
where $I_{45}$ is the neutron star's moment of inertia in units of $10^{45}$
g-cm$^2$.  The numerical value is obtained from the observed \citep{D25}
spin frequency $\omega$ and derivative $\dot\omega$.

FRB 20201124A is identified \citep{F21} with a galaxy at $z = 0.0979$, a
distance of 400 Mpc.  At this distance, the power radiated by a source
with flux density $F_{Jy}$ in Janskys into a solid angle $\Omega$ over the
500 MHz observing bandwidth is
\begin{equation}
	\label{rad}
	P_{rad} \approx 1 \times 10^{41} F_{Jy} {\Omega \over 4\pi} \
	\text{erg/s}.
\end{equation}

Defining $\epsilon$ as the efficiency of converting spindown power to
radiation in the observing bandwidth, Eqs.~\ref{spindown} and \ref{rad} 
yield
\begin{equation}
	\label{Omega}
	\Omega \approx 4 \pi \left({I_{45} \over F_{Jy}}\right)
	\left({\epsilon \over 0.01} \right) \times 2 \times 10^{-7}\ 
	\text{sterad}.
\end{equation}

Known radio pulsars have $10^{-6} \lesssim \epsilon \lesssim 10^{-2}$.  The
brightest bursts of FRB 20201124A have $F_{Jy}$ up to 10 \citep{D25}.  Taking
$I_{45} = 1$ and $F_{Jy} = 10$,
\begin{equation}
	\Omega \approx 2.5 \times 10^{-7}\left({\epsilon \over 0.01}\right)\ 
	\text{sterad}.
\end{equation}
For a conical beam of half-angle $\theta$
\begin{equation}
	\label{theta}
	\theta \approx \sqrt{\Omega \over \pi} \approx 3 \times 10^{-4}
	\sqrt{\epsilon \over 0.01}\ \text{rad}.
\end{equation}
This sets a lower bound on the Lorentz factor of the radiating charges
\begin{equation}
	\Gamma \gtrsim \theta^{-1} \sim 3 \times 10^3\
	\sqrt{0.01 \over \epsilon}.
\end{equation}
For a fan beam the upper bound on $\theta$ may be as small as $\Omega$.
\citet{K18,K20} set other lower bounds on $\Gamma$ by requiring
that it be large enough that the radiating charge bunches are not disrupted
by their Coulomb repulsion.
\section{Emission Region}
The radiated beam is broadened beyond the intrinsic radiation pattern of
accelerated charges if it is emitted by charges with imperfectly collimated
velocity vectors. 
For curvature radiation in a neutron star's magnetosphere the dimension
$\delta r$ of the radiating region $\delta r \lesssim \theta r$.  The
radiation energy contained within a region of volume $(\delta r)^3$ is
$E_{rad} = P_{rad}(\delta r/c)$ (this is much less that the energy of the
burst if its width exceeds $\delta r/c \sim 10\,$ns).  The energy density
\begin{equation}
	\label{Edensity}
	{\cal E}_{rad} = {E_{rad} \over (\delta r)^3} \sim 2 \times 10^{32}
	\left({\epsilon \over 0.01}\right)^{-1/2} \left({I_{45} \over
	F_{Jy}}\right)^{-1/2}\ \text{erg/cm}^3.
\end{equation}
If the factors in parentheses are unity, ${\cal E}_{rad}$ equals the stress
of a magnetic field of $7 \times 10^{16}$ Gauss, and cannot be contained by
plausible magnetar fields.

A small subset of magnetic field lines are not (in a magnetostatic
approximation good to a fractional accuracy $\sim \omega r/c \sim 1 \times
10^{-4}$) bent: for example, those along the axis of a dipole field.  Along
that axis the radiation may be emitted by region of length $\sim 0.3 r$,
multiplying the result of Eq.~\ref{Edensity} by a factor $\delta r/(0.3 r)
\sim 10^{-3}$, reducing the corresponding magnetic field to $\sim 2 \times
10^{15}$ Gauss, consistent with the polar field of a magnetar dipole and the
spindown rate of FRB 20201124A.  Higher multipole fields may be larger
because they are effectively unconstrained by the spindown rate.
\section{Discussion}
Comparing the spindown rate and flux of FRB 20201124A estimates the
collimation of its radiation and bounds the Lorentz factor of its emitting
charges.  It also suggests that the radiation is emitted along open field
lines, as previously suggested by \citet{BK25}, because their large radii of
curvature imply a comparatively large volume can contribute to collimated
radiation, relaxing the constraint on the magnetic field that confines the
radiating charges.

Alignment of the magnetic dipole and rotation axes of active repeating FRB,
as previously suggested \citep{L25}, would explain the difficulty of finding
their rotational periodicity because radiation from a field accurately
azimuthally symmetric around the rotation axis would not be rotationally
modulated.  It is implausible that neutron star fields are accurately
dipolar near the surface, so this argument suggests that FRB are emitted at
many neutron star radii.  Emission along open field lines is consistent
with, but does not require, that the periodic activity modulation of some
repeating FRB results from free precession of the neutron star \citep{ZL20}.

FRB emission along the magnetic dipole axis may also explain the distinction
between repeating and non-repeating FRB:  If aligned magnetic and rotational
axes point in our direction, we may observe all the source's bursts as a
repeating FRB with intrinsic duty factor DF.  If the magnetic and rotational
axes are misaligned we can only observe bursts when, fortuitously and
infrequently, the magnetic axis points toward us.  We can observe only a
fraction ${\cal O}(\Omega/4\pi)$ of all active repeaters close and powerful
enough for detection because the others are beamed away from us
\citep{BK25}.  We can observe a large fraction of sufficiently close and
powerful mis-aligned sources, but only as apparent non-repeaters with duty
factors ${\cal O} (\text{DF}\Omega/4\pi)$; approximating the magnetic axes
of the apparent non-repeaters as uniformly distributed on the sky,
$\Omega/4\pi$ is the fraction of the time they are pointed in our direction.
This is consistent with constraints \citep{K24b} on their duty factors, but
both $\Omega$ and DF are uncertain even to order of magnitude.

If the preceding is valid, then the ratio in the local Universe of aligned
to mis-aligned rotators is roughly the ratio of repeaters to non-repeaters
in observations (such as those of CHIME-FRB) with a wide acceptance angle,
or uncorrelated with the directions of known repeaters, provided the
cumulative duration of the observation exceeds the intervals between bursts
of repeaters.  In practice, this is complicated by the facts that these
intervals depend on observing sensitivity, likely vary among repeating
sources and that the distribution of intervals between bursts is not
Poissonian \citep{K24c}.

The phenomenological differences \citep{P21} between bursts from repeaters
and from apparent non-repeaters, such as the ``sad trombone'' effect, may
then be attributed to differences in the electrodynamics of aligned and
mis-aligned rotators.  This is consistent with the assumption that FRB are
powered, like pulsars, by spin-down energy, not magnetostatic energy.
\section*{Data Availability}
This theoretical study produced no new data.

\end{document}